\documentclass[twocolumn]{revtex4}
\usepackage{amsmath}
\usepackage{graphicx}
\usepackage{epstopdf}
\usepackage{hyperref}
\usepackage{subfigure,xcolor}
\usepackage{braket}
\usepackage{txfonts}

\newcommand{\bea}{\begin{eqnarray}}
\newcommand{\eea}{\end{eqnarray}}
\newcommand{\beq}{\begin{equation}}
\newcommand{\eeq}{\end{equation}}
\newcommand{\nn}{\nonumber}

\def\/{\over}

\begin{document}
\title{\bf Boundarylike behaviors of the resonance interatomic energy in a cosmic string spacetime}
\author{Wenting Zhou$^{1,2}$ and Hongwei Yu$^{2,3}$~\footnote{Corresponding author: hwyu@hunnu.edu.cn}}
\affiliation{$^{1}$Center for Nonlinear Science and Department of Physics, Ningbo
University, Ningbo, Zhejiang 315211, China\\
$^{2}$China Key Laboratory of Low Dimensional Quantum Structures and Quantum Control of Ministry of Education, Hunan Normal University, Changsha,
Hunan 410081, People’s Republic of China\\
$^{3}$Department of physics, Synergetic Innovation Center for Quantum Effects and Applications, Hunan Normal University, Changsha, Hunan 410081, China}

\begin{abstract}
By generalizing the formalism proposed by Dalibard, Dupont-Roc and Cohen Tannoudji, we study the resonance interatomic energy of two identical atoms coupled to quantum massless scalar fields in a symmetric/antisymmetric entangled state  in the Minkowski  and cosmic string spacetimes. We find that in both spacetimes, the resonance interatomic energy has  nothing to do with the field fluctuations but is attributed to the radiation reaction of the atoms only.  We then concretely calculate the resonance interatomic energy of two static atoms near a perfectly reflecting boundary in the Minkowski spacetime and near an infinite and straight cosmic string respectively. We show that the resonance interatomic energy in both cases can be enhanced or suppressed and even nullified as compared with that in an unbounded Minkowski spacetime, because of the presence of the boundary in the Minkowski spacetime or the nontrivial spacetime topological structure of the cosmic string. Besides, we also discover that the resonance interatomic energy in the cosmic string spacetime exhibits some peculiar properties, making it in principle possible to sense different cosmic string spacetimes via the resonance interatomic energy.
\end{abstract}

\maketitle

\section{INTRODUCTION}
Radiative properties of atoms are not inherent as they are crucially dependent on the quantum fields to which they are coupled. For example, for atoms  in cavities or near a perfectly reflecting boundary, the atomic transition rates~\cite{PT82,Wu99,Meschede90,Arias16}, energy shifts~\cite{Meschede90,Wu00,Rizzuto07,Rizzuto091,Rizzuto09,Zhu09,Zhu10,She10,Rizzuto11}, interatomic interaction energy~\cite{PT82,Spagnolo,Passante07} and so on differ from those in a free space. The differences originate from the fact that the modes of quantum fields are modified due to the presence of boundaries. The aforementioned  examples are concerned with the atomic radiative properties in a Minkowski spacetime. In recent years, investigations concerning the atomic radiative properties in curved spacetimes have been carried out~\cite{Iliadakis,Yu071,Yu072,Zhou101,Zhou102,Mello,Zhang13,Zhou16,Menezes171,Menezes172,Zhou171}, and it is found that in certain curved spacetimes,  behaviors of the atomic radiative properties similar to those in a bounded Minkowski spacetime occur, even when boundaries are absent.

The cosmic string spacetime is one such spacetime. The appearance of cosmic strings is one of the predictions of various models of grand unification theories (GUTs). They are extended, one-dimensional, closed and infinite linear objects with a linear mass density and a linear tension that arise as topological defects during the symmetry breaking phase transitions in the early universe~\cite{Kibble,Vilenkin}. The existence of them presumably can induce various astrophysical effects. For example, they are thought to be probable sources of gravitational waves~\cite{Damour,Brandenberger,Jackson}, gamma-ray bursts~\cite{Berezinsky} and high-energy cosmic rays~\cite{Brandenberger09}. Though observational data of the cosmic microwave background has ruled out the cosmic strings as a candidate  for the formation of galaxies, the interest in this topic reappeared in the context of the ``brane-world" scenarios of the superstring theory~\cite{Dvali,Dvali03,Nakano,Ellis,Gergely,Copeland}. Due to its important role in cosmology, there is growing interest in the studies on  cosmic strings in recent years.

The simplest  cosmic string spacetime is that of a static and straight-line cosmic string. It looks like a direct product of the two-dimensional Minkowski space and a cone. Outside the string, the spacetime is locally flat but topologically nontrivial. Fields propagating in such a spacetime are influenced  by its nontrivial topology, and the energy densities of various fields are found to exhibit boundary-induced effects~\cite{Helliwell,Linet,Allen,Frolov,Dowker,Allen92,Davies}. For particles interacting with quantum fields in such a spacetime, though there is no local gravitational force exerting on them, the particles are ``interacting'' with the global conical structure. Such an interaction causes modifications in the radiative properties of particles. It has been shown that the transition rates of atoms in interaction with quantum scalar or electromagnetic fields are affected by the presence of a cosmic string~\cite{Iliadakis,Zhou16,Bilge,Davies,Amirkhanjan}, and a polarizable microparticle around an infinite straight cosmic string senses a Casimir-Polder force~\cite{Saharian}. In this paper, we study another atomic radiative property, i.e., the resonance interatomic energy of two correlated atoms, in the cosmic string spacetime.

Resonance interaction  occurs when two atoms/molecules prepared in different eigenstates are coupled to quantum fields, two atoms with one in the ground state and the other in an excited state, for example. As the exchange of real photons can be involved in such interactions, the interaction energy generally manifests peculiar properties in contrast to the dispersion energy between two atoms in their ground states. For two neutral atoms in their ground states and in interaction with the quantum electromagnetic field, Casimir and Polder showed in their pioneering work that the interaction energy decays with the power law $R^{-7}$ (with $R$ being the interatomic separation) in the long distance region (distances much larger than the typical transition wavelength of atoms)~\cite{Casimir48}. The interaction energy between one atom in the ground state and the other in the excited state, however, scales as $R^{-2}$~\cite{McLone65,Gomberoff66,Power93,Power931,Power95,Rizzuto04,Sherkunov07,Preto13,Donaire15,Milonni15,Berman15,Jentschura17}~\footnote{This behavior can be monotonic or sinusoidally modulated. The monotonic behavior is associated with an irreversible rate of excitation from the initially excited atom to the initially unexcited atom, while the sinusoidally modulated behavior is associated with the reversible, back-and-forth excitation exchange between the two atoms. For details, see Refs.~\cite{Donaire15,Milonni15,Berman15,Jentschura17}.}.
Treatments on these two types of interaction energy entail a fourth-order perturbation theory. But when two identical atoms are prepared in a correlated state, a symmetrical/antisymmetrical entangled state for example, the resonance interaction energy turns out to be a second-order effect. It oscillates with the interatomic separation with the amplitude at large interatomic separation decaying with the power law $R^{-1}$.

The aforementioned properties are of the resonance interaction energy for two atoms in interaction with the quantum electromagnetic field in a free Minkowski spacetime. The resonance interaction energy may be significantly modified if the situations differ. A presence of boundaries is one of such circumstances.
It has been shown that the presence of a conducting plate induces modifications in the interatomic potential of two atoms in their ground states~\cite{Power82,Spagnolo,Passante07}, the near-surface effect sufficiently changes the resonance interaction between an excited atom and a ground-state atom~\cite{Yuan}, and suitably arranged structured environments can significantly affect the resonance energy transfer process~\cite{Andrews15,Andrews16,Andrews17}.
These works refer to two uncorrelated atoms or molecules. Recently, it is also found that the resonance interaction of two correlated atoms exhibits peculiar properties in different environments. For example, the resonant interatomic force between two identical atoms in a symmetric/antisymmetrical entangled state can be greatly enhanced or suppressed in a photonic crystal~\cite{Incardone,Notararigo}, and the transition rate of such two atoms near perfect mirrors may undergo a reduction or an enhancement~\cite{Svaiter16}. Very recently, we studied the resonance interaction energy between two static atoms interacting with quantum electromagnetic fields near a perfectly reflecting boundary and correlated by an entangled state. We discovered that for some specific geometric configurations of the two-atom system with respect to the mirror, the resonance interaction energy exhibits new behaviors as compared with those of two static ones in an unbounded Minkowski spacetime~\cite{Lattuca,Zhou172}.

The cosmic string spacetime is characterized by its nontrivial topological structure, and many quantum effects such as the vacuum fluctuations~\cite{Helliwell,Linet,Allen,Frolov,Dowker,Allen92,Davies}, atomic transition rates\cite{Iliadakis,Zhou16,Bilge,Davies,Amirkhanjan} and energy shifts~\cite{Saharian} in this spacetime exhibit behaviors similar to those in a bounded flat spacetime. Here, we are interested in how the resonance interatomic energy is affected by the presence of a cosmic string and whether some boundarylike effects show up in the resonance interatomic energy. With this  in mind, we study the resonance interaction energy between two atoms in a symmetric/antisymmetric entangled state and in interaction with a quantum massless scalar field near a perfectly reflecting boundary in the Minkowski spacetime and near an infinite and  straight cosmic string respectively. By comparing the results in these two different backgrounds, we also explore the possibility of sensing the cosmic string via the resonance interatomic energy.

The paper is organized as follows. In Sec. II, we introduce the formalism proposed by Dalibard, Dupont-Roc and Cohen-Tannandji (DDC formalism)~\cite{DDC82,DDC84}, and by which the resonance interatomic energy will be calculated. In Secs. III and IV, we use the DDC formalism to evaluate the resonance interatomic energy of two identical two-level atoms prepared in the symmetric/antisymmetric entangled state and in interaction with quantum massless scalar fields in two different backgrounds, i.e., near a perfectly reflecting boundary in the Minkowski spacetime and in the cosmic string spacetime. Finally, we give conclusions in Sec. V.

\section{The DDC formalism}
We assume that two two-level identical atoms labeled by $A$ and $B$ are in interaction with quantum massless scalar fields in a Minkowski/cosmic string spacetime. The field operators in these two spacetimes can be expressed in the following general form:
\beq
\phi(t,\vec{x})=\sum_{\vec{k}}[a_{\vec{k}}(t)u_{\vec{k}}(\vec{x})+\text{H.c.}]\;,
\eeq
in which $a^{\dag}_{\vec{k}}$ is the creation operator of the field,  and ``H.c.'' denotes the Hermitian conjugate. The specific expressions of the basic modes, $u_{\vec{k}}(\vec{x})$,  will be given in Sec. III for the scalar field in a bounded Minkowski spacetime and in Sec. IV for the scalar field in a cosmic string spacetime, as they are closely dependent on the concrete configurations of the spacetimes.

The atoms possess two stationary eigenstates $|+\rangle$ and $|-\rangle$ with energies $\pm{\omega_0\/2}$, and are assumed to be static. Then the Hamiltonian of the system composed by the two atoms and the scalar field is given by
\bea
H&=&\omega_0{\sigma}^A_3(\tau)+\omega_0{\sigma}^B_3(\tau)+\lambda ({\sigma}^A_2(\tau)\phi(x_A(\tau))\nn\\&&
+{\sigma}^B_2(\tau)\phi(x_B(\tau)))+\sum_{\vec{k}}\omega_{\vec{k}}a^{\dag}_{\vec{k}} a_{\vec{k}}{dt\/d\tau}\;,
\eea
where $\tau$ is the atomic proper time~\footnote{It is worth pointing out here that for the case of two static atoms, $t=\tau$. However, as the formalism we generalize in the paper is also valid for other cases such as two atoms in uniform  acceleration  with a constant interatomic separation, in which the atomic coordinate time doesn't coincide with the proper time, we express the operators in terms of the atomic proper time $\tau$, in order to keep the generality.}, $\lambda$ is the coupling constant, and
\bea
{\sigma}^{\xi}_2(\tau)&=&{i\/2}({\sigma}^{\xi}_-(\tau)-{\sigma}^{\xi}_+(\tau))\;,\\
{\sigma}_3(\tau)&=&{1\/2}(|+\rangle\langle+|-|-\rangle\langle-|)
\eea
with $\xi=A,B$, ${\sigma}_-(\tau)=|-\rangle\langle+|$ and ${\sigma}_+(\tau)=|+\rangle\langle-|$.

For atoms interacting with quantum fields, the atomic radiative properties may be attributed to vacuum fluctuations~\cite{Welton,Compagno} and radiation reaction~\cite{Ackerhalt}, or a combination of them~\cite{Milonni,Milonni88}. The ambiguity arises from the ordering of operators of fields and atoms.  Dalibard, Dupont-Roc and Cohen Tannoudji(DDC) proposed that a symmetric ordering should be exploited, so that the contributions of field fluctuations and radiation reaction can be distinctively separated~\cite{DDC82,DDC84}. This formalism has been widely used to study the spontaneous transition rates and energy shifts of a single atom interacting with quantum fields in various environments~\cite{Meschede90,Rizzuto07,Rizzuto091,Rizzuto09,Zhu09,Zhu10,She10,Rizzuto11,Iliadakis,Yu071,Yu072,Zhou101,Zhou102,Zhou16,Menezes171,Menezes172}, and very recently, it was generalized to study the resonance interaction between two atoms prepared in an entangled state in the Minkowski spacetime~\cite{Zhou172,Rizzuto16,Zhou161,Arias16,Menezes16,Menezes161} and in the Schwarzschild spacetime~\cite{Zhou171,Menezes16}. Here we introduce the DDC formalism with which we will investigate the resonance interaction energy of two atoms in the Minkowski and cosmic string spacetimes.

To evaluate the resonance interatomic energy of the two atoms, we choose to work in the Heisenberg picture. We firstly write out the Heisenberg equations of motion for the dynamical variables of the field and the atoms respectively, and then we solve these equations and divide each solution into a free part and a source part which are denoted by the superscript ``f'' and ``s'' respectively. For the dynamical variable of the field, we have $a_{\vec{k}}(t(\tau))=a^f_{\vec{k}}(t(\tau))+a^s_{\vec{k}}(t(\tau))$ with
\beq
a^f_{\vec{k}}(t(\tau))=a_{\vec{k}}(t(\tau_0))e^{-i\omega_{\vec{k}}(t(\tau)-t(\tau_0))}
\eeq
and
\bea
a^s_{\vec{k}}(t(\tau))&=&i\lambda\int^{\tau}_{\tau_0}d\tau'{\sigma}^{Af}_2(\tau')[\phi^f(x_A(\tau')),a^f_{\vec{k}}(t(\tau))]\nn\\
&+&i\lambda\int^{\tau}_{\tau_0}d\tau'{\sigma}^{Bf}_2(\tau')[\phi^f(x_B(\tau')),a^f_{\vec{k}}(t(\tau))]\;.
\eea
Then the field operator can be accordingly expressed as: $\phi(t(\tau),\vec{x}(\tau))=\phi^f(t(\tau),\vec{x}(\tau))+\phi^s(t(\tau),\vec{x}(\tau))$ with
\beq
\phi^f(t(\tau),\vec{x}(\tau))=\sum_{\vec{k}}[a^f_{\vec{k}}(t)u_{\vec{k}}(\vec{x})+\text{H.c.}]
\eeq
and
\bea
\phi^s(t(\tau),\vec{x}(\tau))&=&i\lambda\int^{\tau}_{\tau_0}d\tau'{\sigma}^{Af}_2(\tau')[\phi^f(x_A(\tau')),\phi^f(x(\tau))]\nn\\
&+&i\lambda\int^{\tau}_{\tau_0}d\tau'{\sigma}^{Bf}_2(\tau')[\phi^f(x_B(\tau')),\phi^f(x(\tau))]\;.\nn\\
\eea
Similar treatments on the dynamical variables of the atoms give rise to the following free part and source part for the operators ${\sigma}^{\xi}_2(\tau)$:
\bea
{\sigma}^{\xi f}_2(\tau)&=&{i\/2}[{\sigma}^{\xi}_{-}(\tau_0)e^{- i\omega_0(\tau-\tau_0)}-{\sigma}^{\xi}_{+}(\tau_0)e^{i\omega_0(\tau-\tau_0)}]\;,\\
{\sigma}^{\xi s}_2(\tau)&=&i\lambda\int^{\tau}_{\tau_0}d\tau'[{\sigma}^{\xi f}_2(\tau'),{\sigma}^{\xi f}_2(\tau)]\phi^f(x_{\xi}(\tau'))\;,
\eea
and those for the operators ${\sigma}^{\xi}_3(\tau)$:
\beq
{\sigma}^{\xi f}_3(\tau)={\sigma}^{\xi}_3(\tau_0)\;,
\eeq
\beq
{\sigma}^{\xi s}_3(\tau)=i\lambda\int^{\tau}_{\tau_0}d\tau'[{\sigma}^{\xi f}_2(\tau'),{\sigma}^{\xi f}_3(\tau)]\phi^f(x_{\xi}(\tau'))\;.
\eeq
Notice that all the above source parts are accurate to the first order of the coupling constant.

Using the above free parts and source parts of the atomic and field's operators in the Heisenberg equations of motion, and choosing a symmetric ordering between the operators of the atoms and the field, we can distinguish the contributions of field fluctuations and radiation reaction to the variation rate of  the atomic energy. For atom $A$, the expectation value of the contribution of field fluctuations over the vacuum state of the field is given by
\bea
\biggl({d H_A(\tau)\/d\tau}\biggr)_{vf}&=&-\lambda^2\omega_0\int^{\tau}_{\tau_0}d\tau'[{\sigma}^{Af}_2(\tau'),
[{\sigma}^{Af}_2(\tau),{\sigma}_3^{Af}(\tau)]]\nn\\
&&\times C^F(x_A(\tau),x_A(\tau'))
\eea
with $C^F(x_A(\tau),x_A(\tau'))$ being the symmetric correlation function of the field defined as
\beq
C^F(x_A(\tau),x_A(\tau'))={1\/2}\langle0|\{\phi^f(x_A(\tau)),\phi^f(x_A(\tau'))\}|0\rangle\;,
\eeq
and that of the radiation reaction is
\bea
\biggl({d H_A(\tau)\/d\tau}\biggr)_{rr}&=&\lambda^2\omega_0\int^{\tau}_{\tau_0}d\tau'
\{[{\sigma}^{Af}_2(\tau),{\sigma}^{Af}_3(\tau)],{\sigma}^{Af}_2(\tau')]\}\nn\\
&&\times\chi^F(x_A(\tau),x_A(\tau'))\nn\\
&+&\lambda^2\omega_0\int^{\tau}_{\tau_0}d\tau'\{[{\sigma}^{Af}_2(\tau),{\sigma}^{Af}_3(\tau)],{\sigma}_2^{Bf}(\tau')\}\nn\\
&&\times\chi^F(x_A(\tau),x_B(\tau'))
\eea
with $\chi^F(x_A(\tau),x_B(\tau'))$ being the linear susceptibility function of the field defined as
\beq
\chi^F(x_A(\tau),x_B(\tau'))={1\/2}\langle0|[\phi^f(x_A(\tau)),\phi^f(x_B(\tau'))]|0\rangle\;.\label{chif}
\eeq
From the above expressions of $({d H_A(\tau)\/d\tau})_{vf,rr}$, we can identify the effective Hamiltonians of the contributions of vacuum fluctuations to atom $A$:
\bea
H^{eff}_{A,vf}(\tau)&=&{1\/2}i\lambda^2\int^{\tau}_{\tau_0}d\tau'[{\sigma}^{Af}_2(\tau'),{\sigma}_2^{Af}(\tau)]\nn\\
&&\quad\times C^F(x_A(\tau),x_A(\tau'))\;,\label{vf-eff-A}
\eea
and that of the radiation reaction to atom $A$:
\bea
H^{eff}_{A,rr}(\tau)&=&-{1\/2}i\lambda^2\int^{\tau}_{\tau_0}d\tau'\{{\sigma}^{Af}_2(\tau),{\sigma}_2^{Bf}(\tau')\}\nn\\
&&\quad\times\chi^F(x_A(\tau),x_B(\tau'))\nn\\
&&-{1\/2}i\lambda^2\int^{\tau}_{\tau_0}d\tau'\{{\sigma}^{Af}_2(\tau),{\sigma}_2^{Af}(\tau')\}\nn\\
&&\quad\times\chi^F(x_A(\tau),x_A(\tau'))\;.\label{rr-eff-A}
\eea
Equation (\ref{vf-eff-A}) shows that the effective Hamiltonian for the contribution of vacuum fluctuations to atom $A$ is independent of atom $B$, and thus it has no contribution to the interatomic energy. However, in Eq. (\ref{rr-eff-A}), which is the effective Hamiltonian of the radiation reaction, though the second term only depends on atom $A$, the first term depends on both atoms, and thus it contributes to the interatomic energy. Taking similar steps for atom $B$, we can get $H^{eff}_{B,vf}(\tau)$ and $H^{eff}_{B,rr}(\tau)$.

As mentioned previously, the two atoms are prepared in the symmetric/antisymmetric entangled state:
\beq
|\psi_{\pm}\rangle={1\/\sqrt{2}}(|g_A e_B\rangle\pm|e_A g_B\rangle),
\eeq
in which ``$g$'' and ``$e$'' represent the ground and excited states of the atoms.

Then the summation of the average values of the two-atom dependent parts in both $H^{eff}_{A,rr}(\tau)$ and $H^{eff}_{B,rr}(\tau)$ over the above states gives rise to the following formula for the resonance interatomic energy:
\bea
\delta E&=&-i\lambda^2\int^{\tau}_{\tau_0}d\tau'C^{AB}(\tau,\tau')\chi^F(x_A(\tau),x_B(\tau'))\nn\\&&
+A\rightleftharpoons B\;term\label{formula-RIE}
\eea
with
\beq
C^{AB}(\tau,\tau')={1\/2}\langle\psi_{\pm}|\{{\sigma}^{Af}_2(\tau),{\sigma}^{Bf}_2(\tau')\}|\psi_{\pm}\rangle\;,
\eeq
which is the symmetric statistical function of the atoms. For two two-level identical atoms, this function can be further simplified to be
\beq
C^{AB}(\tau,\tau')=\pm{1\/8}(e^{i\omega_0(\tau-\tau')}+e^{-i\omega_0(\tau-\tau')})\label{two-level-CAB}
\eeq
with ``$\pm$'' corresponding to atoms in states $|\pm\rangle$ respectively. From the above derivation, we see that the resonance interatomic energy between two atoms in a symmetric/antisymmetric entangled state is independent of the vacuum fluctuations of fields, and results only from the  radiation reaction of atoms.

In the following two sections, we use this formalism to calculate the resonance interatomic energy of two identical static atoms in the symmetric/antisymmetric entangled state in the Minkowski  and cosmic string spacetimes respectively.

\section{Resonance interatomic energy of two static atoms near a perfectly reflecting boundary in the Minkowski spacetime}

\begin{figure}[!htbp]
\centering
\includegraphics[width=0.30\textwidth]{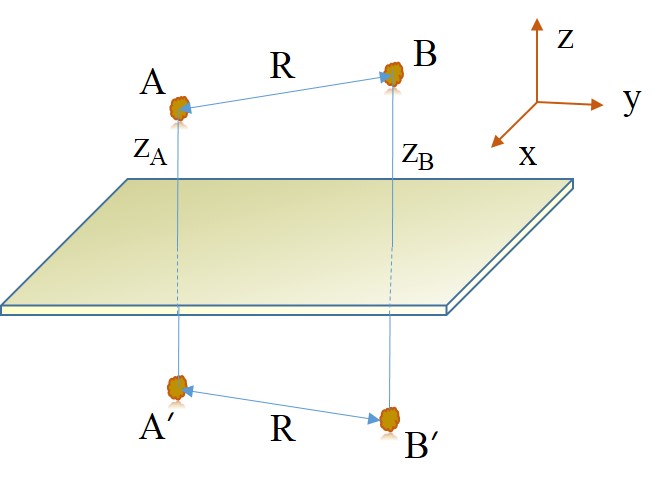}
\caption{Two static atoms with a separation $R$ are located near a perfectly reflecting boundary in the Minkowski vacuum. $A'$ and $B'$ are the images of atoms $A$ and $B$ respectively.}
\label{fig:Normal-position-Min}
\end{figure}

As shown in Fig.~\ref{fig:Normal-position-Min}, we suppose that two identical two-level atoms with a separation $R$ and prepared in the symmetric/antisymmetric entangled state are located near a perfectly reflecting boundary in the Minkowski vacuum, and they are in interaction with a quantum massless scalar field. For this case, we choose the rectangular coordinate system to study the resonance interatomic energy, and assume that the boundary coincides with the ``$xoy$" plane.

The field operator near the boundary can be described by
\beq
\phi(t,\vec{x})=\int_{k_z>0} d^3\vec{k}[a_{\vec{k}}u_{\vec{k}}(t,\vec{x})+H.c.]
\eeq
%\beq
%\phi(t,\vec{x})={1\/2\pi}\sqrt{2\/\pi}\int_{k_z>0} {d^3\vec{k}\/\sqrt{2\omega}}[a_{\vec{k}}e^{-i\omega t}e^{i\vec{k}_{\|}\cdot\vec{x}_{\|}}\sin({k_z z})+H.c.]
%\eeq
with
\beq
u_{\vec{k}}(t,\vec{x})={1\/2\pi\sqrt{\pi\omega}}e^{-i\omega t}e^{i\vec{k}_{\|}\cdot\vec{x}_{\|}}\sin({k_z z})\;,
\eeq
$\vec{k}_{\|}=(k_x,k_y)$, and $\vec{x}_{\|}=(x,y)$. It is easy to check that as $k_z,k_z^{'}>0$, the above modes satisfy the following relation:
\bea
&&(u_{\vec{k}}(t,\vec{x}),u_{\vec{k'}}(t,\vec{x}))\nn\\
&=&-i\int d^3x\;(u_{\vec{k}}(t,\vec{x})\partial_t u^\ast_{\vec{k'}}(t,\vec{x})
-\partial_t u_{\vec{k}}(t,\vec{x})u^\ast_{\vec{k'}}(t,\vec{x}))\nn\\&=&\delta_{\vec{k}\vec{k'}}\;.
\label{product}
\eea
Then the two-point correlation function of the field follows:
\bea
&&\langle0|\phi(t_A,\vec{x}_A)\phi(t_B,\vec{x}_B)|0\rangle
=-{1\/4\pi^2}{1\/{(t_A-t_B-i\epsilon)^2-R^2}}\nn\\&&\quad\quad\quad\quad\quad\quad+
{1\/4\pi^2}{1\/{(t_A-t_B-i\epsilon)^2-\bar{R}^2}}
\label{two-point-corre-boundary}
\eea
with
\bea
R^2&=&(x_A-x_B)^2+(y_A-y_B)^2+(z_A-z_B)^2\;,\\
\bar{R}^2&=&(x_A-x_B)^2+(y_A-y_B)^2+(z_A+z_B)^2\;.
\eea
Notice that the second term in the two-point correlation function, Eq.~(\ref{two-point-corre-boundary}), reflects the effect of image atoms $A'$ and $B'$, and it can  be derived by using the method of images.

\begin{figure}[!htbp]
\centering
\includegraphics[width=0.35\textwidth]{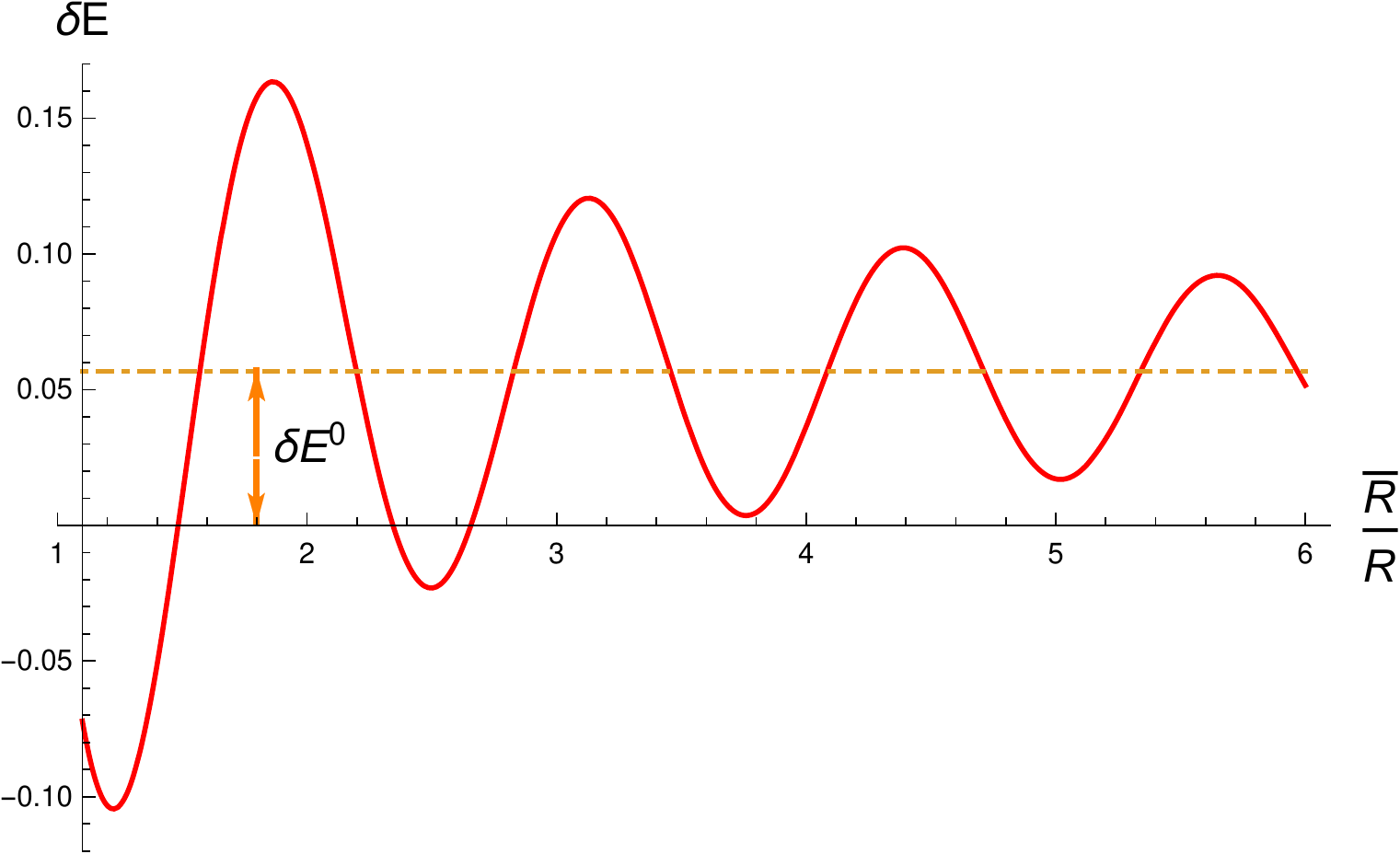}
\caption{The location[with respect to the boundary]-dependence of the resonance interatomic energy of two atoms with a fixed interatomic separation near the boundary. The atomic transition frequency and the interatomic separation are respectively $\omega_0=1.549\times10^{16}s^{-1}$ and $R=9.674\times10^{-8}m$.
The ordinate is of unit $\mp{\lambda^2\omega_0\/16\pi}$.}
\label{Fig0}
\end{figure}

Combining Eq.~(\ref{two-point-corre-boundary}) with Eq.~(\ref{chif}), we obtain the following linear susceptibility function of the field:
\bea
&&\chi^F(x_A(\tau),x_B(\tau'))\nn\\
&=&-{1\/8\pi^2}\biggl[{1\/{(\Delta\tau-i\epsilon)^2-R^2}}-{1\/{(\Delta\tau+i\epsilon)^2-R^2}}\biggr]\nn\\&&
+{1\/8\pi^2}\biggl[{1\/{(\Delta\tau-i\epsilon)^2-\bar{R}^2}}-{1\/{(\Delta\tau+i\epsilon)^2-\bar{R}^2}}\biggr]\;.\;\;\;\;
\eea
Using the above function together with (\ref{two-level-CAB}) in Eq.~(\ref{formula-RIE}), and doing some calculations with the technique of contour integration and the residuum theory, we get the following resonance interatomic energy for the two atoms near the boundary:
\beq
\delta E=\mp{\lambda^2\/16\pi}\biggl({\cos(\omega_0 R)\/R}-{\cos(\omega_0 \bar{R})\/\bar{R}}\biggr)\;.\label{Min-boundary}
\eeq
This result shows that the resonance interatomic energy of the two atoms near the boundary is composed by two parts: the first part is only dependent on  the relative positions of the two atoms  and it is completely the same as the resonance interatomic energy of two static atoms in an unbounded Minkowski spacetime [see the result obtained by taking $a\rightarrow0$ in Eq.~(14) of Ref.~\cite{Rizzuto16}]; the second part however depends on locations of the two atoms with respect to the boundary, and thus it results from the presence of the boundary.  In general, the interaction can either be strengthened or weakened depending on the positions of the atoms. In particular, for some special locations of the two atoms with respect to the boundary, which satisfies the relation, $\bar{R}\cos(\omega_0 R)\leq R\cos(\omega_0\bar{R})$, the second term equals or  even overwhelms the first term.  As a result, the resonance interatomic energy of the two atoms can be dramatically manipulated by the presence of the boundary as compared with that in an unbounded space [denoted by $\delta E^0$], being nullified or even causing a sign change of the interaction.
This means that the effect of the atomic radiation reaction on the resonance interatomic energy can be completely shielded by the presence of the boundary for atoms at some particular locations while at some other locations the presence of the boundary alters the sign of the interaction. For a graphic example, see Fig.~\ref{Fig0}.

To show more clearly the effect of the presence of a perfectly reflecting boundary on the resonance interatomic energy of the two atoms in the symmetric/antisymmetric entangled state, we
compare the resonance interatomic energy of two atoms near the boundary in two different orientations with respect to it  [denoted by $\delta E^{\parallel}$ and $\delta E^{\perp}$] with that in an unbounded space [$\delta E^{0}$].
\begin{figure}[!htbp]
\centering
\subfigure[]{
\includegraphics[width=0.35\textwidth]{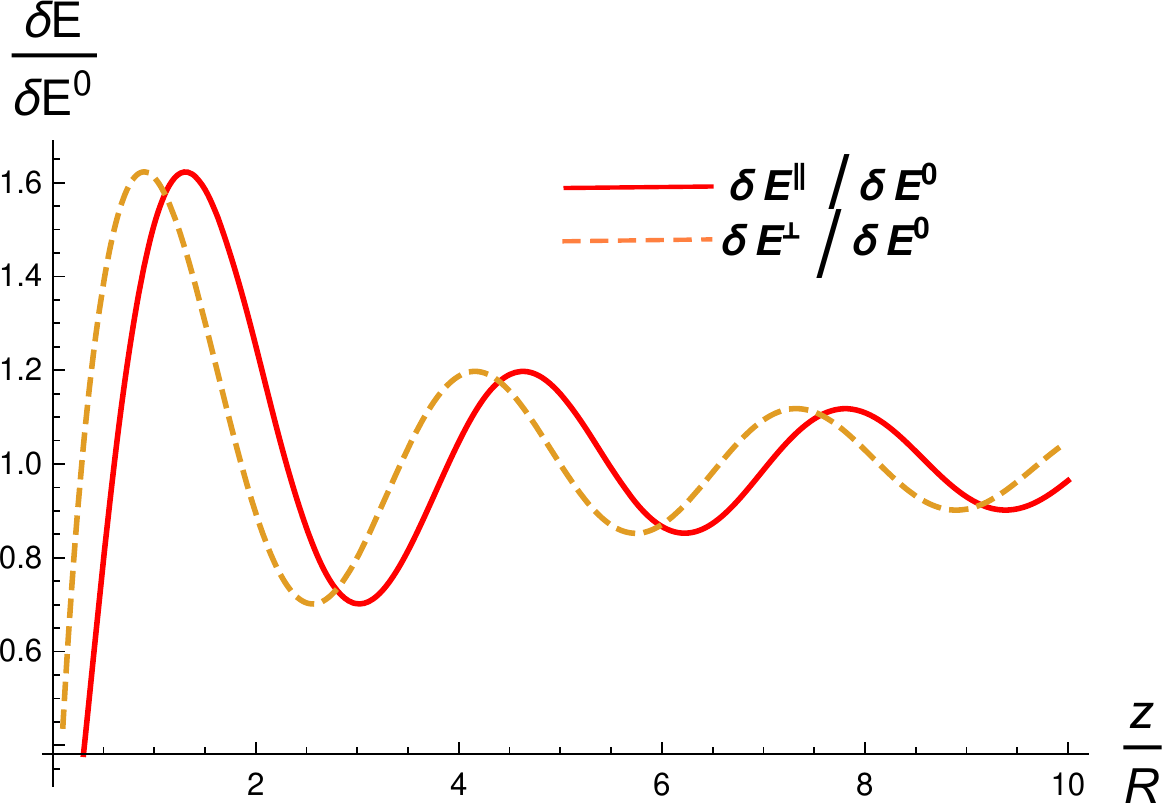}\label{Fig1.pdf}}
\subfigure[]{
\includegraphics[width=0.35\textwidth]{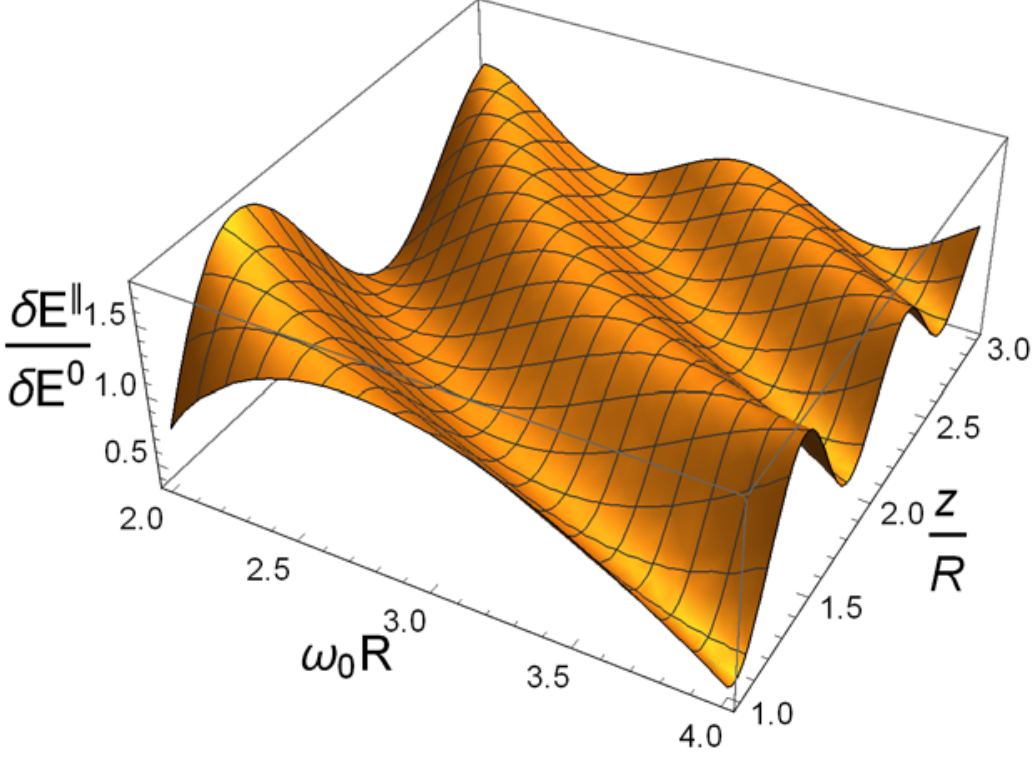}\label{Fig2.pdf}}
\caption{The relative resonance interatomic energy of two static atoms near a perfectly reflecting boundary. (a) Two specific configurations are considered, i.e., two atoms aligned with their separation parallel or perpendicular to the boundary. For the former case, $z_A=z_B=z$, and the resonance interaction energy is represented by $\delta E^{\parallel}$; while for the latter case, $z$ represents the distance between the boundary and the atom closer to the boundary, and the resonance interaction energy is represented by $\delta E^{\perp}$. Parameters are chosen such that $\omega_0R=1$. (b) The relative resonance interatomic energy of two atoms aligned with the interatomic separation parallel to the boundary and with various values of interatomic separation and atom-boundary separation. For two atoms located with their separation perpendicular to the plate, $\delta E^{\perp}\/\delta E^0$ exhibits similar behaviors as $\delta E^{\parallel}\/\delta E^0$ shown in this figure.}
\label{boundary-RIE-N}
\end{figure}
%In Fig.~\ref{Fig1}, \textcolor{red}{both the resonance interatomic energy of two atoms located with their separation parallel or perpendicular to the boundary, relative to its counterpart in an unbounded space, i.e. $\delta E\/\delta E_0$, are plotted.} % For two atoms located with the interatomic separation parallel to the boundary, $z=z_A=z_B$; while for two atoms located with the interatomic separation perpendicular to the boundary, $z$ denotes the separation between the boundary and the atom which is closer to the boundary.}
Figures.~\ref{Fig1.pdf} and \ref{Fig2.pdf} show that the relative resonance interatomic energy of the two atoms with constant separation $R$ near the boundary oscillates with the atom-plate separation $z$ around unity, where $z$ is the distance between the boundary and the atom which is closer. It means that the resonance interatomic energy near the boundary can be larger or smaller than  that in an unbounded space, depending on the relative positions of the two atoms with respect to the plate. As the effect of the boundary becomes weaker with increasing atom-plate separation, the oscillation amplitude decreases with increasing atom-plate separation. As a result, the resonance interatomic energy  approaches, at infinity, the corresponding value in an unbounded Minkowski spacetime. %\textcolor{red}{Fig.~\ref{Fig5} shows the resonance interatomic energy of two atoms  as a function  of the interatomic separations perpendicular to the boundary. The distance between the boundary and the atom closer to the boundary is chosen to be constant $z=0.5\omega_0^{-1}$, and $R$ ranges from $0.2z$ to $20z$. It's obvious that the resonance interatomic energy in this region is smaller than that in an unbounded space. With the increasing of the interatomic separation, this interatomic energy becomes smaller and smaller, and almost vanishes as $R\sim\bar{R}$ when $z/R\gg1$.}

\section{Resonance interatomic energy of two static atoms in a cosmic string spacetime}

\begin{figure}[!htbp]
\centering
\includegraphics[width=0.20\textwidth]{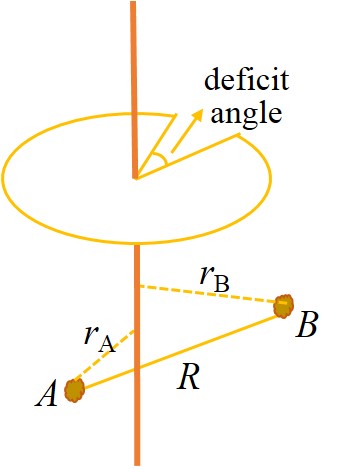}
\caption{Two atoms with separation $R$ are located near an infinite and straight cosmic string.}
\label{cs-general-position}
\end{figure}

In this section, we study the resonance interatomic energy of two identical two-level atoms prepared in the symmetric/antisymmetric entangled state near a static, straight, and infinitely long and thin cosmic string[see Fig.~\ref{cs-general-position}].

Suppose that the string lies along the $z$ axis, then we can describe its spacetime metric in the cylindrical coordinates $(t,r,\theta,z)$ as
\beq
ds^2=dt^2-dr^2-r^2d\theta^2-dz^2\;,\label{cs metric}
\eeq
where $0\leq\theta<{2\pi/\nu}$, $\nu=(1-4G\mu)^{-1}$ with $G$ and $\mu$ being the Newton constant and the mass per unit length of the string. This metric corresponds to a very simple exact solution of the Einstein equations. The value of $\nu$ is determined by the value of the mass density of the string which is in turn determined by the spontaneous symmetry breaking scale when the cosmic string was formed. For typical GUT models, $G\mu\sim10^{-6}$, which gives rise to a value of $\nu$ slightly larger than unity. The existence of the cosmic string does not produce any local gravitational field, but it does induce a nontrivial global topology, in the sense that a length of a unit circle around the string is less than $2\pi$ with a deficit angle $8\pi G\mu$ [as shown in Fig.~\ref{cs-general-position}], and a surface of constant $(t,z)$ exhibits the geometry of a cone rather than that of a plane.

In this spacetime, the Klein-Gordon equation of the massless scalar field takes the form
\beq
\partial^2_t-{1\/r}\partial_r(r\partial_r)-{1\/r^2}\partial^2_{\theta}-\partial^2_z\phi(t,\vec{x})=0\;.
\eeq
Solving this equation by a separation of variables, we can get a complete set of normalized field modes~\cite{Iliadakis}:
\beq
u_{m,\kappa,k_{\perp}}(t,\vec{x})={1\/2\pi}\sqrt{\nu\/2\omega}e^{-i\omega t}e^{i\kappa z}e^{i\nu m\theta}J_{\nu|m|}(k_{\perp}r)\;,\label{mode cs}
\eeq
with $\kappa\in(-\infty,\infty)$, $m\in Z$, $k_{\perp}\in[0,\infty)$, $\omega^2=\kappa^2+k^2_{\perp}$ and $J_{\nu|m|}(k_{\perp}r)$ being the first kind Bessel function.
The product [defined in the second line of Eq.~(\ref{product})] of these modes follows as
\beq
(u_{m,\kappa,k_{\perp}}(t,\vec{x}),u_{m^{'},\kappa^{'},k^{'}_{\perp}}(t,\vec{x}))=
\delta(\kappa-\kappa')\delta_{mm'}{\delta(k_{\perp}-k^{'}_{\perp})\/\sqrt{k_{\perp}k^{'}_{\perp}}}\;.
\eeq
Then we can expand the massless scalar field in terms of the above modes as
\beq
\phi(t,\vec{x})=\int d\mu_j[a_j u_j(t,\vec{x})+\text{H.c.}]\label{cs-field-operator}
\eeq
with $j\equiv\{m,\kappa,k_{\perp}\}$ and
\beq
\int d\mu_j=\sum^{\infty}_{m=-\infty}\int^{\infty}_{-\infty}d\kappa\int^{\infty}_0dk_{\perp}k_{\perp}\;.
\eeq
From the mode function, Eq.~(\ref{mode cs}), we can easily deduce that the field operator in the cosmic string spacetime is periodic [with respect to the coordinate $\theta$]  with a periodicity ${2\pi\/\nu}$  rather than $2\pi$.

With the field operator (\ref{cs-field-operator}), we can easily obtain the linear susceptibility function of the field,
\bea
&&\chi^F(x_A(\tau),x_B(\tau'))={\nu\/16\pi^2}\sum^{\infty}_{m=-\infty}\int^{\infty}_{-\infty}d\kappa\int^{\infty}_0dk_{\perp}{k_{\perp}\/\omega}e^{i\kappa z}\nn\\&&
\quad\times e^{i\nu m\Delta\theta}J_{\nu|m|}(k_{\perp}r_A)J_{\nu|m|}(k_{\perp}r_B)(e^{-i\omega\Delta\tau}-e^{i\omega\Delta\tau})
\eea
with $\Delta\tau=\tau-\tau'$, $z=z_A-z_B$ and $\Delta\theta=\theta_A-\theta_B$.

Substituting this function and the symmetric statistical function of the atoms, Eq.~(\ref{two-level-CAB}), into Eq.~(\ref{formula-RIE}), and doing some simplifications, we arrive at
\bea
&&\delta E=\mp{\lambda^2\nu\/32\pi^2}\sum_{m=-\infty}^{\infty}\int^{\infty}_{-\infty}d\kappa\int^{\infty}_0dk_{\perp}{k_{\perp}\/\omega}e^{i\kappa z}e^{i\nu m\Delta\theta}\nn\\&&
\quad\times J_{\nu|m|}(k_{\perp}r_A)J_{\nu|m|}(k_{\perp}r_B)\biggl({1\/\omega+\omega_0}+{1\/\omega-\omega_0}\biggr)\;,\quad\quad\label{general-RIE}
\eea
which is the general expression of the resonance interatomic energy of two identical static atoms in a symmetric/antisymmetric entangled state and in interaction with the quantum massless scalar field in the cosmic string spacetime. As $\kappa^2+k^2_{\perp}=\omega^2$, by making transformations $k_{\perp}=\omega\sin\varphi$ and $\kappa=\omega\cos\varphi$, the above formula is accordingly transformed into
\bea
\delta E&=&\mp{\lambda^2\nu\/16\pi^2}\sum_{m=-\infty}^{\infty} e^{i\nu m \Delta\theta}\int^{\pi\/2}_{0}d\varphi\int^{\infty}_0d\omega\omega\sin\varphi\nn\\
&&\times\cos(\omega z\cos\varphi)J_{\nu|m|}(\omega r_A\sin\varphi)J_{\nu|m|}(\omega r_B\sin\varphi)\nn\\
&&\times\biggl({1\/\omega+\omega_0}+{1\/\omega-\omega_0}\biggr)\;.\label{general-RIE-1}
\eea
For general positions of the two atoms with respect to the string in a cosmic string spacetime with noninteger parameter $\nu$, further simplifications of the above expression is very difficult. However, for some special cases, we can still derive some analytical results.

(1)$\nu=1$.

Using the property of the first kind Bessel function that
\beq
\sum^{\infty}_{m=-\infty}e^{i m\Delta\theta}J_{|m|}(k_{\perp}r_A)J_{|m|}(k_{\perp}r_B)=J_{0}(k_{\perp}\mathcal{R})
\eeq
with $\mathcal{R}=\sqrt{r^2_A+r^2_B-2r_Ar_B\cos\Delta\theta}$\;, the resonance interatomic energy, Eq.~(\ref{general-RIE-1}), reduces to
\bea
\delta E&=&\mp{\lambda^2\/16\pi^2}\int^{\pi\/2}_{0}d\varphi\int^{\infty}_0d\omega\omega\sin\varphi \cos(\omega z\cos\varphi)\nn\\
&&\times J_{0}(\omega\mathcal{R}\sin\varphi)\biggl({1\/\omega+\omega_0}+{1\/\omega-\omega_0}\biggr)\;.
\eea
Further simplifications of the above integration gives rise to
\beq
\delta E=\mp{\lambda^2\/16\pi}{\cos(\omega_0R)\/R}\;,\label{Min-RIE-unbounded}
\eeq
with $R=\sqrt{\mathcal{R}^2+z^2}$, which is exactly the interatomic separation. The above resonance interatomic energy is the same as the first term in Eq.~(\ref{Min-boundary}), which is the resonance interatomic energy of two atoms in an unbounded Minkowski spacetime. The consistency comes out naturally, as when $\nu=1$, the deficit angle disappears and the cosmic string spacetime reduces to the Minkowski spacetime.

(2)$\nu>1$.

(2.1) One or both atoms are located on the string.

We firstly consider the simplest case in which at least one of the two atoms is located on the string, i.e., $r_A=0$ or $r_B=0$. Notice that
\beq
J_{\alpha}(0)=\left\{
                \begin{array}{ll}
                  0, & \alpha>0\;, \\
                  1, & \alpha=0\;,
                \end{array}
              \right.\label{Jalpha-0}
\eeq
then if  both atoms  are located on the string, the resonance interatomic energy, Eq.~(\ref{general-RIE-1}), can be simplified to be:
\bea
\delta E&=&\mp{\lambda^2\nu\/16\pi^2}\int^{\pi\/2}_{0}d\varphi\int^{\infty}_0d\omega\omega\sin\varphi \cos(\omega z\cos\varphi)\nn\\
&&\times\biggl({1\/\omega+\omega_0}+{1\/\omega-\omega_0}\biggr)\nn\\
&=&\mp{\lambda^2\nu\/16\pi}{\cos(\omega_0z)\/|z|}\;;\label{RIE-both-atoms-on-string}
\eea
for the case in which only one atom is located on the string (atom $A$ for example), the use of relation (\ref{Jalpha-0}) in Eq.~(\ref{general-RIE-1}) leads to:
\bea
\delta E&=&\mp{\lambda^2\nu\/16\pi^2}\int^{\pi\/2}_{0}d\varphi\int^{\infty}_0d\omega\omega\sin\varphi \cos(\omega z\cos\varphi)\nn\\
&&\times J_0(\omega r_B\sin\varphi)\biggl({1\/\omega+\omega_0}+{1\/\omega-\omega_0}\biggr)\nn\\
&=&\mp{\lambda^2\nu\/16\pi}{\cos(\omega_0\sqrt{r_B^2+z^2})\/\sqrt{r_B^2+z^2}}\;.\label{RIE-one-atom-on-string}
\eea
Notice that if both atoms are located on the string, $|z|$ is equal to the interatomic separation, and for the case in which only atom $A$ is located on the string, $\sqrt{r_B^2+z^2}$ also represents the interatomic separation. Thus when one or both atoms are located on the string, we have
\beq
\delta E=\mp{\lambda^2\nu\/16\pi}{\cos(\omega_0R)\/R}\;.\label{RIE-atleastone-atom-on-string}
\eeq
Compare it with its counterpart in an unbounded Minkowski spacetime [see Eq.~(\ref{Min-RIE-unbounded})],  we find that they are quite similar, except for the presence of the parameter $\nu$ in the above result. As $\nu>1$, it reveals that for the case where one or both atoms  are located on the string, the resonance interatomic energy of two identical atoms in a symmetrical/antisymmetrical  entangled state is amplified to be $\nu$ times  that in an unbounded Minkowski spacetime.

(2.2) Two atoms fixed in a cosmic string spacetime with integer $\nu$.

Though the parameter $\nu$ for a real cosmic string spacetime is only slightly larger than unity, the investigations on cosmic string spacetimes with integer $\nu$ also caught much  attention~\cite{Zhou16,Davies,Bilge,Amirkhanjan,Saharian,Mota}, as for this case analytical results are usually obtainable, and they are very helpful for us to understand the properties of the cosmic string spacetimes.

For integer $\nu$, using the following property of the first kind Bessel function~\cite{Davies},
\bea
{1\/\nu}\sum^{\nu-1}_{n=0}J_0(k L_{n,\nu})&=&2\sum^{\infty}_{m=1}J_{\nu m}(k r_A)J_{\nu m}(k r_B)\cos(\nu m\Delta\theta)\nn\\&&
+J_0(k r_A)J_0(k r_B)
\eea
with $L_{n,\nu}=\sqrt{r^2_A+r^2_B-2r_Ar_B\cos(\Delta\theta+2\pi n/\nu)}$, in the general expression of the resonance interaction energy [Eq.~(\ref{general-RIE-1})], we get
\bea
\delta E&=&\mp{\lambda^2\/16\pi^2}\sum_{n=0}^{\nu-1}\int^{\infty}_0d\omega\omega\biggl({1\/\omega+\omega_0}+{1\/\omega-\omega_0}\biggr)\nn\\&&
\times\int^{\pi\/2}_0d\varphi\sin\varphi\cos(\omega z\cos\varphi)J_0(\omega L_{n,\nu}\sin\varphi)\;,\quad\nn\\
&=&\mp{\lambda^2\/16\pi}\sum^{\nu-1}_{n=0}{\cos(\omega_0R_{n,\nu})\/R_{n,\nu}}\label{integer nu-RIE}
\eea
with $R_{n,\nu}=\sqrt{z^2+L_{n,\nu}^2}$. Notice that when $r_A\sim0$ or $r_B\sim0$, or both $r_{A,B}\sim0$, this result approaches Eq.~(\ref{RIE-atleastone-atom-on-string}).

For $\nu=1$, the above result reduces to
\beq
\delta E=\mp{\lambda^2\/16\pi}{\cos(\omega_0R_{0,1})\/R_{0,1}}
\eeq
with $R_{0,1}=\sqrt{z^2+r^2_A+r^2_B-2r_Ar_B\cos\Delta\theta}=R$, which is exactly the resonance interatomic energy of two atoms in an unbounded Minkowski spacetime. So by taking $\nu=1$, we recover the resonance interatomic energy of two atoms in a free Minkowski space.

For $\nu=2$, Eq.~(\ref{integer nu-RIE}) can be reexpressed to be a sum of two terms:
\beq
\delta E=\mp{\lambda^2\/16\pi}\biggl[{\cos(\omega_0R_{0,2})\/R_{0,2}}+{\cos(\omega_0R_{1,2})\/R_{1,2}}\biggr]
\eeq
with
\bea
&&\quad\quad\quad\quad\;\; R_{0,2}=R_{0,1}=R\;,\\
&&R_{1,2}=\sqrt{z^2+r^2_A+r^2_B+2r_Ar_B\cos\Delta\theta}\;.
\eea
Notice that the first term is only determined by the interatomic separation between the two atoms, and it is completely  the same as the resonance interatomic energy of two atoms in a free Minkowski vacuum; while the second term is closely dependent on the locations of the two atoms with respect to the string. Comparing it with the resonance interatomic energy of two static atoms located near a perfectly reflecting boundary in the Minkowski vacuum [see Eq.~(\ref{Min-boundary})], we find that they are quite similar. However, for the resonance interatomic energy of two atoms near the perfectly reflecting boundary, the second term in Eq.~(\ref{Min-boundary}) is induced by the presence of the boundary; while the second term in the above result appears in a cosmic string spacetime. For cosmic string spacetimes with integer $\nu\geq3$, more such terms show up.

The following figures graphically  show the effect of the presence of a straight cosmic string on the resonance interatomic energy of two static atoms in an entangled state. We take the case of two atoms aligned with their separation parallel to the string for an example.
\begin{figure}[!htbp]
\centering
\subfigure[]{
\includegraphics[width=0.40\textwidth]{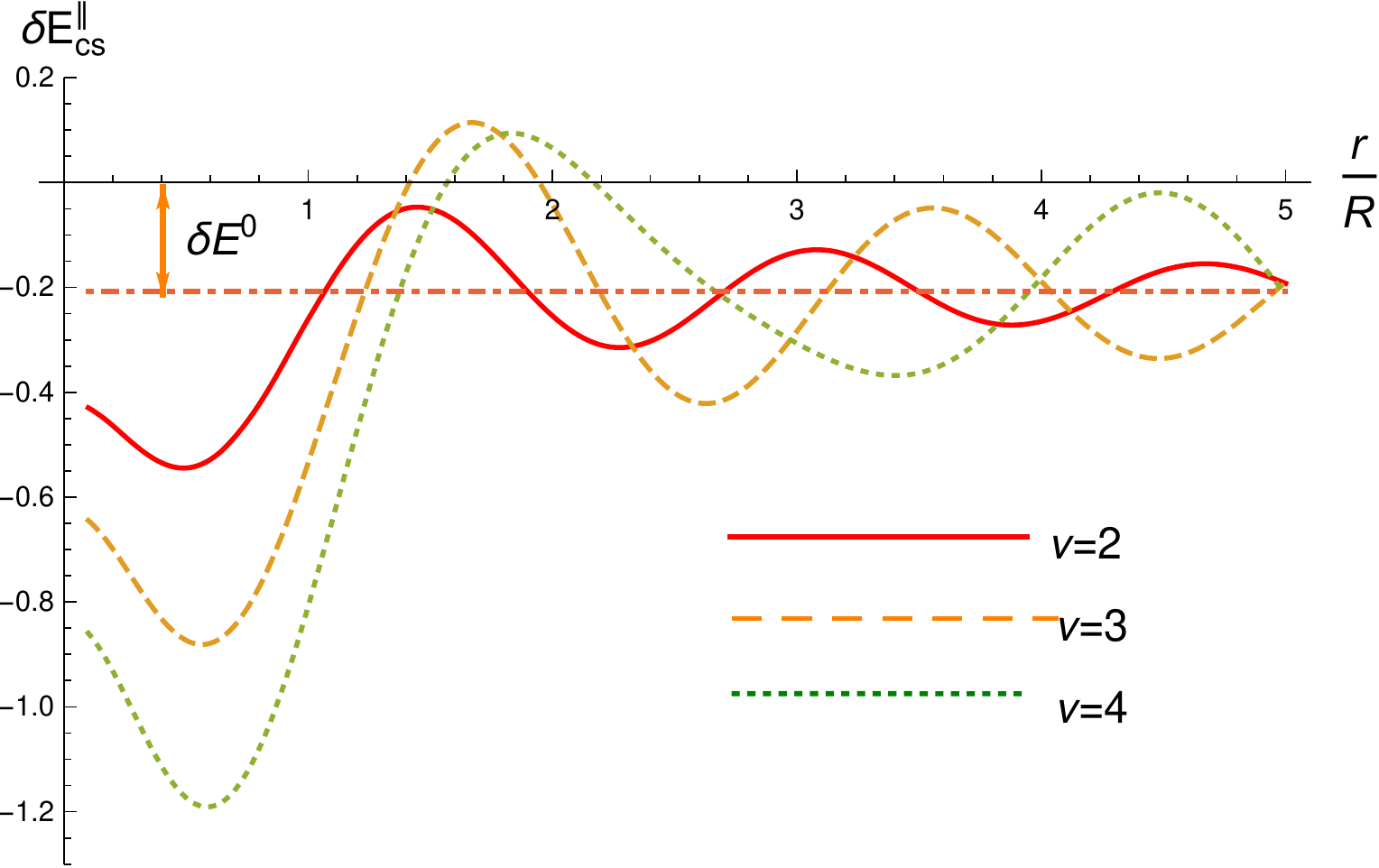}\label{Fig5.pdf}
\label{Fig5}}
\subfigure[]{
\includegraphics[width=0.35\textwidth]{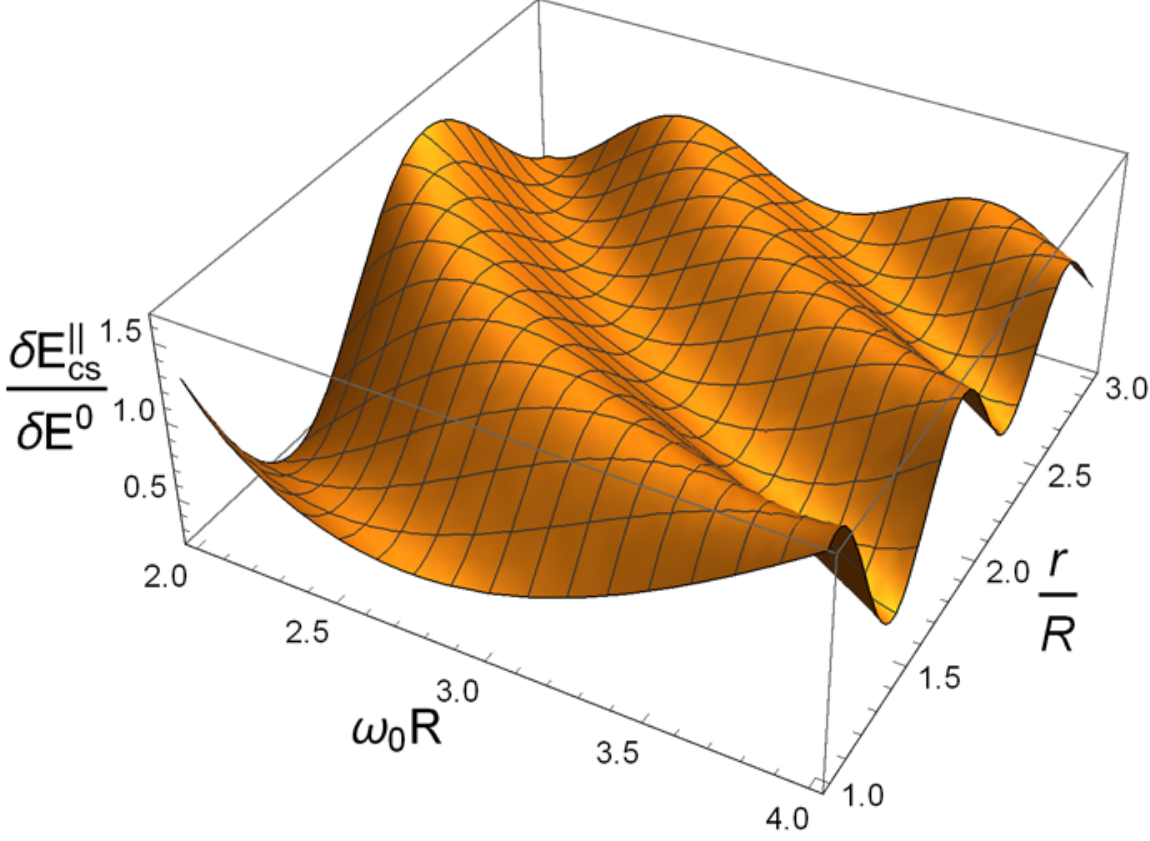}\label{Fig4.pdf}}
\caption{The resonance interatomic energy of two atoms fixed with constant separation $R$ parallel to an infinite and straight cosmic string. We denote the atom-string separation for both atoms by $r$, i.e., $r_A=r_B=r$. (a) The resonance interatomic energy of two atoms in cosmic string spacetimes with different values of the parameter $\nu$. Parameters are chosen such that $\omega_0R=2$. The ordinate is of unit $\mp{\lambda^2\omega_0\/16\pi}$. (b) The interatomic-separation and atom-string-separation dependence of the relative resonance interatomic energy of two atoms in a cosmic string spacetime with $\nu=2$.}
\label{parallel-cs}
\end{figure}
As shown in Figs.~\ref{Fig5.pdf} and \ref{Fig4.pdf}, the resonance interatomic energy of the two atoms shows behaviors very much like those of two atoms located near a perfectly reflecting boundary in the Minkowski spacetime. It oscillates with the atom-string separation, with the oscillation amplitude decreasing with increasing atom-string distance. Thus when $r\rightarrow\infty$, the resonance interatomic energy approaches the corresponding value in an unbounded Minkowski spacetime. The solid red, dashed yellow and dotted green lines in Fig.~\ref{Fig5.pdf} depict the atom-string-separation dependence of the resonance interatomic energy of the two atoms with the same interatomic separation $R=\omega_0^{-1}$ in the cosmic string spacetimes with various values of $\nu$. Notice that for atoms in the cosmic string spacetimes with  proper values of $\nu$, the dashed yellow and dotted green lines corresponding to the cosmic string spacetimes with $\nu=3,4$ for example, the resonance interatomic energy can vanish or change sign at certain positions near the string, revealing that the effect of the atomic radiation reaction on the resonance interatomic energy can be drastically changed by the presence of a cosmic string. This is very similar to the case of two atoms near a perfectly reflecting boundary in the Minkowski spacetime, in which the effect of atomic radiation reaction on the resonance interatomic energy can, for example, also be completely screened at certain locations by the presence of a perfectly reflecting boundary. These similarities can be ascribed to the peculiar properties of the cosmic string spacetime, i.e., it is characterized by a deficit angle, and the spacetime topology is nontrivial. As such a spacetime is only locally but not globally flat, field modes propagating in the spacetime are ``restricted" by the special structure, thus atoms interacting with the quantum fields in the spacetime exhibit  behaviors similar to those in a bounded Minkowski spacetime. Despite the similarities, the resonance interatomic energy in a cosmic string spacetime also distinguishes  itself with peculiar properties. We discover from Fig.~\ref{Fig5.pdf} that when the atoms-string separation is much shorter than the interatomic separation, i.e. ${r\/R}\ll1$, the value of the resonance interatomic energy is almost $\nu$ times  that in an unbounded Minkowski spacetime. This conclusion is in consistence with the analytical result, Eq.~(\ref{RIE-atleastone-atom-on-string}). For different values of the parameter $\nu$, the behaviors of the resonance interatomic energy are distinctive. Particularly, the oscillation of the resonance interatomic energy with the atom-string separation is more severe in a cosmic string spacetime with larger parameter $\nu$. Thus principally speaking, it is probable to identify different cosmic string spacetimes via the resonance interatomic energy.

\section{CONCLUSIONS}
We generalized the DDC formalism for studying the resonance interatomic energy of two identical static atoms in a symmetric/antisymmetric entangled state and in interacting with quantum scalar fields in the Minkowski spacetime and the cosmic string spacetime. We found that in both spacetimes, the resonance interatomic energy between the two atoms is caused only by the radiation reaction of the two atoms. By exploiting this formalism, we studied the resonance interatomic energy of such two atoms near a perfectly reflecting boundary in the Minkowski spacetime and near an infinite and straight comic string respectively. As compared with that in an unbounded Minkowski spacetime, $\delta E^{0}$, the resonance interatomic energy in the former case is revised by the presence of the boundary. It can be enhanced or suppressed and even nullified depending on the relative locations of the two atoms with respect to the boundary. For the two atoms  in a cosmic string spacetime, if one or both atoms are located on the string, the resonance interatomic energy is $\nu$ times  that in an unbounded Minkowski spacetime [$\delta E^{0}$].  For the case where  both atoms are located outside the string, as compared with $\delta E^{0}$, the resonance interatomic energy is revised by the presence of the string in a manner very similar to that by a perfectly reflecting boundary in a Minkowski spacetime. It can also be enhanced or suppressed, depending on the relative positions of the two atoms with respect to the string. Despite the similarities, we discover that the resonance interatomic energy in the cosmic string spacetime also shows some peculiar properties, making it in principle possible to discern different cosmic string spacetimes via the resonance interatomic energy.

In summary, for two identical static atoms in interaction with quantum massless scalar fields near a perfectly reflecting boundary in the Minkowski spacetime or near an infinite and straight cosmic string, the resonance interatomic energy can be enhanced or suppressed, as compared with that in an unbounded Minkowski spacetime, depending on the relative positions of the two atoms with respect to the boundary or the string. The cause of the similarities is the peculiar characters of the cosmic string spacetime, i.e., it is locally flat but not globally. As field modes propagating in such a spacetime are ``restricted'' by the nontrivial topological structure, atoms interacting with the fields exhibit similar properties as those of atoms near a boundary in the Minkowski spacetime.

\addcontentsline{toc}{chapter}{Acknowledgment}
\begin{acknowledgments}
This work was supported in part by the NSFC under
Grants No. 11690034, No. 11435006, and
No. 11405091; the Research Program of
Ningbo University under Grants No. XYL18027,
the K. C. Wong Magna Fund in Ningbo University,
and the Key Laboratory of Low Dimensional
Quantum Structures and Quantum Control of Ministry of
Education under Grants No. QSQC1525.
\end{acknowledgments}

\end{document}